\documentclass[twocolumn]{aastex62}

\usepackage{amsmath}
\accepted{December 14, 2020}

\shorttitle{M dwarf's atmosphere \& wind}
\shortauthors{Sakaue \& Shibata}

\begin{document}

\title{Nonlinear Alfv\'en Wave Model of Stellar Coronae and Winds from the Sun to M dwarfs}

\correspondingauthor{Takahito Sakaue}
\email{sakaue@kwasan.kyoto-u.ac.jp}


\author{Takahito Sakaue}
\affiliation{Astronomical Observatory, Kyoto University, Japan}

\author{Kazunari Shibata}
\affiliation{Astronomical Observatory, Kyoto University, Japan}






\begin{abstract}
  M dwarf's atmosphere and wind is expected to be highly magnetized. The nonlinear propagation of Alfv\'en wave could play a key role in both heating the stellar atmosphere and driving the stellar wind. Along this Alfv\'en wave scenario, we carried out the one-dimensional compressive magnetohydrodynamic (MHD) simulation about the nonlinear propagation of Alfv\'en wave from the M dwarf's photosphere, chromosphere to the corona and interplanetary space. Based on the simulation results, we develop the semi-empirical method describing the solar and M dwarf's coronal temperature, stellar wind velocity, and wind's mass loss rate. We find that M dwarfs' coronae tend to be cooler than solar corona, and that M dwarfs' stellar winds would be characterized with faster velocity and much smaller mass loss rate compared to those of the solar wind.

\end{abstract}


\section{Introduction} \label{sec:intro}

M-type main sequence stars (M dwarfs) have the highly magnetized atmosphere. Their magnetic activities have been particularly discussed with the focus on their impact on the planetary atmosphere \citep{2007AsBio...7...85S,2007AsBio...7...30T,2007AsBio...7..167K,2007AsBio...7..185L,2013Sci...340..577S}. It is important for studies about the exoplanets or astrobiology to reveal the underlying physics for the structure of stellar atmosphere and wind \citep{2014MNRAS.438.1162V,2020MNRAS.494.1297M,2019LNP...955.....L}.

The most promising mechanism for both heating the stellar atmosphere and driving the stellar wind is the nonlinear processes related to the Alfv\'en wave \citep{1993A&A...270..304V,2011ApJ...741...54C}. Alfv\'en wave is responsible for the transfer of magnetic energy in the magnetized plasma and is involved in the energy conversion to the kinetic or thermal energy of the background media through the nonlinear processes. Based on this scenario, the three-dimensional (3D) magnetohydrodynamic (MHD) global model for the solar atmosphere and wind (Alfv\'en Wave Solar Model (AWSoM) by \cite{2014ApJ...782...81V}) has been developed to investigate the stellar wind of M dwarfs and environments around their planets \citep{2014ApJ...790...57C,2017ApJ...843L..33G,2018PNAS..115..260D,2020ApJ...895...47A}.

Although these studies discuss the 3D global structure of stellar wind and magnetic field configuration, their applicability is limited due to the following two properties intrinsic to their models. First, the inner boundary of their models is placed at the ``top of stellar chromosphere'', and the Alfv\'en wave amplitude on that height is given by the empirical law \citep{2013ApJ...764...23S}. Second, the interaction between the Alfv\'en wave and the stellar wind is considered in much simplified way with many analytical, empirical, or phenomenological terms, because the propagating Alfv\'en wave cannot be resolved directly in their 3D simulations owing to the low-spatial resolution. The effect of the other compressible waves on the stellar wind and Alfv\'en wave propagation is neglected.

These difficulties in the above 3D global model have been addressed by the numerical studies about the nonlinear propagation of Alfv\'en wave along the single magnetic flux tube in the solar atmosphere from the photosphere, chromosphere, to the corona \citep{1982SoPh...75...35H,1999ApJ...514..493K,2010ApJ...710.1857M} and solar wind \citep{2005ApJ...632L..49S,2006JGRA..111.6101S,2012ApJ...749....8M,2014MNRAS.440..971M,2018ApJ...853..190S,2019ApJ...880L...2S,2020ApJ...900..120S,2020MNRAS.tmp.3336M}. These approaches also have been extended to the stellar atmosphere and wind models \citep{2013PASJ...65...98S,2018PASJ...70...34S,2020ApJ...896..123S}, and revealed that the Alfv\'en wave amplitude on the top of chromosphere should be self-consistently determined as a consequence of wave dissipation and reflection in the chromosphere. In addition, owing to their high-resolution simulations, it is found that, while the atmosphere and wind are maintained by the energy and momentum transfer by Alfv\'en wave, its propagation is affected by the dynamics of atmosphere and wind. These studies highlight the importance of resolving the local dynamics associated with the Alfv\'en wave propagation, as well as reproducing the global structure of the solar and stellar atmosphere and wind.

In this letter, therefore, we extend our recent solar atmosphere and wind model \citep{2020ApJ...900..120S} to the M dwarf's atmosphere and wind. By carrying out the one-dimensional (1D) time-dependent MHD simulations, the nonlinear propagation of Alfv\'en wave in the nonsteady stellar atmosphere and wind is calculated from the M dwarf's photosphere, chromosphere, to the corona and interplanetary space. The primary goal of this letter is to summarize the differences in the reproduced stellar atmosphere and wind structures between the Sun and M dwarfs. The physical mechanisms for such a diversity of stellar atmosphere and wind are also discussed here and will be more quantitatively investigated in the subsequent paper \citep{subsequent}, in which we develop the semi-empirical method describing the stellar atmosphere and wind parameters (coronal temperature, wind velocity and mass loss rate) based on the simulation results.

\section{NUMERICAL SETTING}

\begin{figure}
  \begin{center}
    \epsscale{.5} 
    \plotone{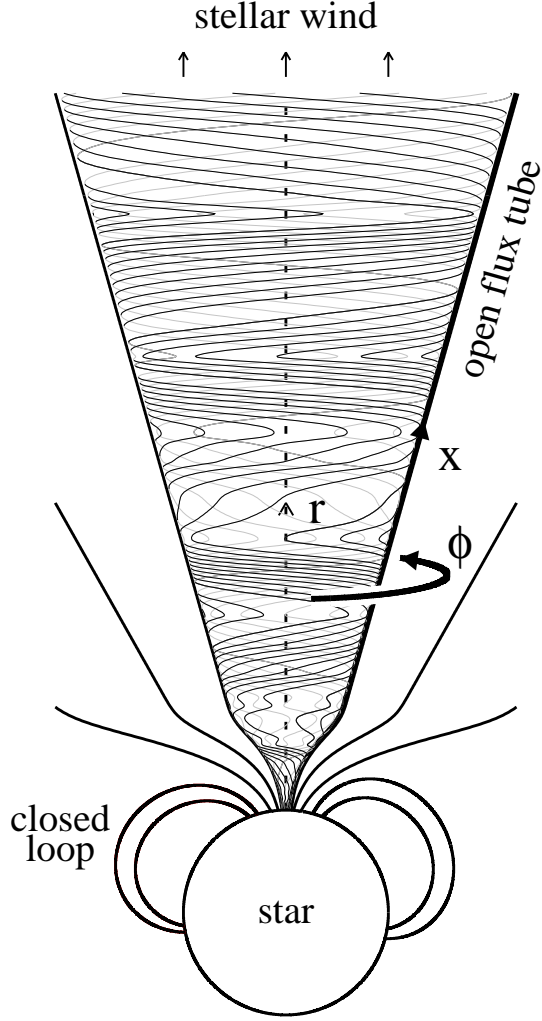}
    \caption{Schematic image of axisymmetric magnetic flux tube, the surface of which is defined by the poloidal $x$ and toroidal $\phi$ axes. The winding thin lines represent the magnetic field lines, which illustrate the nonlinear propagation of Alfv\'en wave.}
    \label{fig:figure01.eps}
  \end{center}
\end{figure}

The nonlinear propagation of the Alfv\'en wave in the time-dependent stellar atmosphere and wind is simulated by using 1D MHD equations based on the axial symmetry assumption of the magnetic flux tube (see Appendix \ref{sec:basic_eq} and \cite{2020ApJ...900..120S}).
The surface of the axisymmetric flux tube is defined by the poloidal and toroidal axes which are noted in this study with $x$ and $\phi$ (Figure \ref{fig:figure01.eps}).
There are three free parameters determining the magnetic flux tube configuration used in this study, including the photospheric magnetic field strength ($B_{\rm ph}$), chromospheric magnetic field strength ($\overline{B}$), filling factor of open flux tube on the photosphere ($f_{\rm ph}$). Among them, $B_{\rm ph}$ is assumed to be equipartition to the photospheric plasma pressure, and $f_{\rm ph}$ is fixed at 1/1600.

By employing the different stellar photospheres as the boundary conditions, we considered the stellar atmospheres and winds of the Sun and two M dwarfs, including AD Leo (M3.5) and TRAPPIST-1 (M8). The stellar mass ($M_\star$), radius ($r_\star$), effective temperature ($T_{\rm eff}$) of AD Leo are $0.47M_\odot$, $0.46r_\odot$, 3473 K, respectively \citep{2015AA...577A.132M}, where $M_\odot=2.0\times10^{33}$ g and $r_\odot=7.0\times10^{10}$ cm are the solar mass and radius. TRAPPIST-1's $M_\star$, $r_\star$, $T_{\rm eff}$ are $0.08M_\odot$, $0.12r_\star$, 2559 K, respectively \citep{2016Natur.533..221G}. These basic parameters imply that M dwarfs are characterized with the larger gravitational acceleration ($g_\star$), shorter pressure scale height of the photosphere ($H_{\rm ph}=RT_{\rm eff}/(\mu_{\rm ph}g_\star)$), and almost the same surface escape velocity $v_{\rm esc\star}$, compared to the Sun. In fact, $\log_{10}g_\star=$4.44, 4.79 and 5.21 for the Sun, AD Leo and TRAPPIST-1, respectively. $H_{\rm ph}=$130, 29, and 6.9 km as well, and $v_{\rm esc\star}=$618, 624, and 511 km s$^{-1}$.

The outwardly propagating Alfv\'en wave is excited on the photosphere by imposing the velocity and magnetic fluctuations on the bottom boundary, which represent the surface convective motion. The mass density and convective velocity on the photosphere is calculated based on the opacity table presented by \cite{2014ApJS..214...25F} and surface convection theory by \cite{1999AA...346..111L,2002AA...395...99L}, and \cite{2015AA...573A..89M}. 
The outer boundary is set at $r\gtrsim100r_\star$, and 19200 grids are placed nonuniformly. The numerical scheme is based on the HLLD Riemann solver \citep{2005JCoPh.208..315M} with the second-order MUSCL interpolation and the third-order TVD Runge-Kutta method \citep{1988JCoPh..77..439S}. The heat conduction is solved by super-time-stepping method \citep{2012MNRAS.422.2102M}.
We also performed the parameter survey about the chromospheric magnetic field strength and the velocity amplitude on the photosphere for each star.

\section{TYPICAL SIMULATION RESULTS}
\label{sec:results_analyses}
\begin{figure}
  \begin{center}
    \epsscale{1.} 
    \plotone{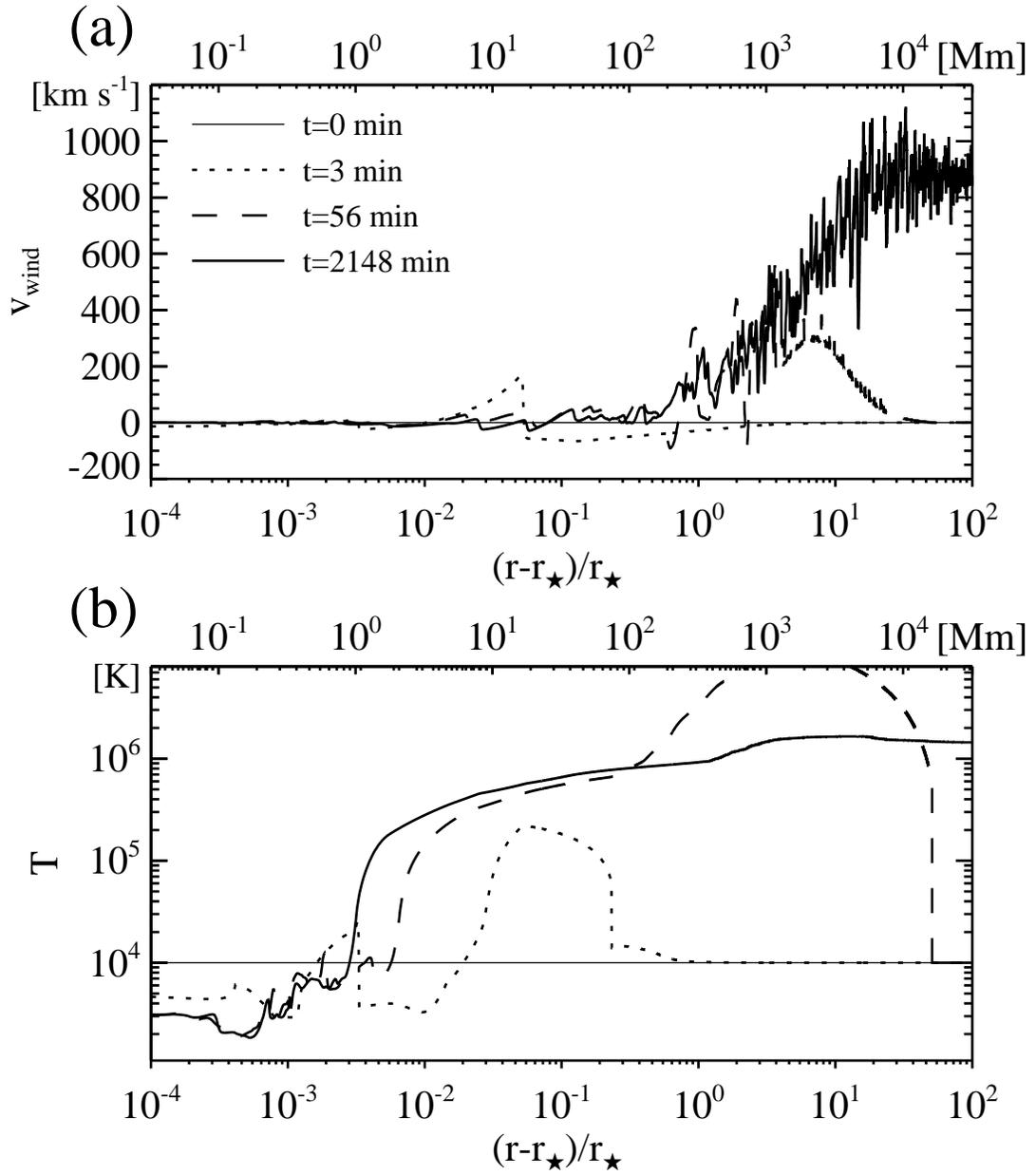}
    \caption{The temporal variations of the stellar wind velocity (panel (a)) and temperature (panel (b)) in the case of M3.5. The stellar wind velocity reaches around 900 km s$^{-1}$. The transition layer appears in the temperature profile, dividing the lower-temperature chromosphere and 10 MK corona.}
    \label{fig:figure02.eps}
  \end{center}
\end{figure}

After several tens of hours, the stellar wind in the simulation box reaches the quasi-steady state. In particular case of M3.5 dwarf shown in Figure \ref{fig:figure02.eps}, it is found that the stellar wind velocity reaches around 900 km s$^{-1}$, and that the transition layer appears in the temperature profile around 1 Mm, dividing the lower-temperature chromosphere and 1 MK corona.

To characterize the physical quantities of quasi-steady state of stellar atmospheres and winds, we investigate the integrals of the basic equations. First is the integral of the equation of motion, which is obtained by temporally averaging and spatially integrating Equation (\ref{eq:mom_pol}).
\begin{align}
  v_x^2(r)=\Delta_p^r+\Delta_{p_B}^r+\Delta_c^r+\Delta_t^r+\Delta_g^r.
  \label{eq:bernoulli}
\end{align}
The right-hand side terms are defined as: $\Delta_p^r=-2\int^r_{r_\star}\left\langle{1\over\rho}{\partial p\over\partial x}\right\rangle dx$, where $\langle\cdot\rangle$ means the temporal average, $\Delta_{p_B}^r=-2\int^r_{r_\star}\left\langle{1\over\rho}{\partial\over\partial x}\left({B_\phi^2\over8\pi}\right)\right\rangle dx$, $\Delta_c^r=2\int^r_{r_\star}\langle v_\phi^2\rangle{\partial\ln\sqrt{A}\over\partial x}dx$, $\Delta_t^r=-2\int^r_{r_\star}\left\langle{B_\phi^2\over4\pi\rho}\right\rangle{\partial\ln\sqrt{A}\over\partial x}dx$, and $\Delta_g^r=-v_{\rm esc\star}^2\left(1-{r_\star\over r}\right)$, where $v_{\rm esc\star}=\sqrt{2GM_\star/r_\star}$ is the escape velocity on the stellar surface.

Another integral of equation describes the energy flux conservation.
\begin{equation}
  A(F_A+F(v_x)+F_g+F_c+F_{\rm rad})=L_{\rm total}=\rm const.
  \label{eq:energy_conservation}
\end{equation}
where $F_A=-B_x\langle B_\phi v_\phi\rangle/(4\pi)$ is Poynting flux by the magnetic tension (Alfv\'en wave energy flux), $F_g=-\langle\rho v_x\rangle GM_\star/r$ is the gravitational energy flux, and $F_{\rm rad}={1\over A}\int^x_\infty A\langle Q_{\rm rad}\rangle dx'$ is the energy flux representing the radiative energy loss. $F(v_x)$ is the sum of enthalpy flux $F_{\rm ent}$, kinetic energy flux $F_{\rm kin}$ and the Poynting flux advected by the stellar wind: $F_{\rm ent}=\gamma\langle pv_x\rangle/(\gamma-1)$, $F_{\rm kin}=\langle\rho v^2v_x\rangle/2$, and $F(v_x)=F_{\rm ent}+F_{\rm kin}+\langle B_\phi^2v_x\rangle/(4\pi)$.

\begin{figure}
  \begin{center}
    \epsscale{.8} 
    \plotone{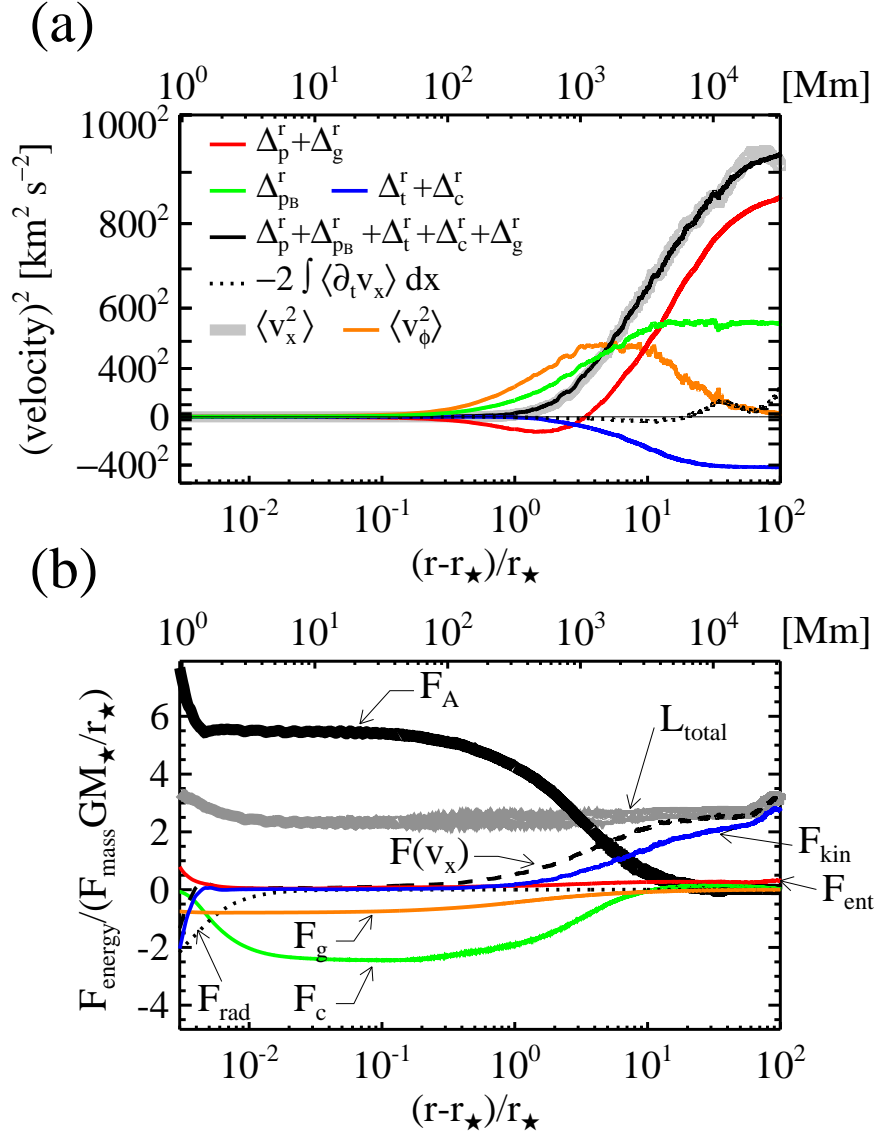}
    \caption{The momentum and energy conservations in the stellar wind of M3.5 dwarf. In panel (a), the profile of $v_x^2$ (thick gray line) is compared to the contributions by magnetic pressure acceleration $\Delta_{p_B}^r$ (green line), sum of plasma pressure $\Delta_p^r$ and gravitational acceleration $\Delta_g^r$ (red line), and sum of centrifugal force $\Delta_c^r$ and magnetic tension force $\Delta_t^r$. The orange line shows the profile of the square amplitude of Alfv\'en wave. Panel (b) shows the energy fluxes normalized by $F_{\rm mass}GM_\star/r_\star$, where $F_{\rm mass}$ is the mass flux and $F_{\rm mass}GM_\star/r_\star\approx5\times10^3r_\star^2/(fr^2)$ erg cm$^{-2}$ s$^{-1}$. $F_A$, $F_{\rm kin}$, $F_{\rm ent}$, $F_g$, $F_{\rm rad}$, $F_c$, and $F(v_x)$ are Alfv\'en wave energy flux, kientic energy flux, enthalpy flux, gravitational energy flux, heat conduction flux, and the sum of enthalpy flux, kinetic energy flux, and Poynting flux advected with the stellar wind. $L_{\rm total}$ is the integral constant in Equation (\ref{eq:energy_conservation}).}
    \label{fig:figure03.eps}
  \end{center}
\end{figure}

Equation (\ref{eq:bernoulli}) and (\ref{eq:energy_conservation}) are confirmed in Figure \ref{fig:figure03.eps}(a) and (b), which show the simulation result of stellar wind of M3.5 dwarf. In Figure \ref{fig:figure03.eps}(a), the black solid line corresponds to $\Delta_p^r+\Delta_g^r+\Delta_{p_B}^r+\Delta_t^r+\Delta_c^r$, which agrees well with $v_x^2$ (thick gray line) as indicated by Equation (\ref{eq:bernoulli}). It is most remarkable in Figure \ref{fig:figure03.eps}(a) that the stellar wind is mainly driven by the plasma pressure gradient (red solid line). In particular, the slow shocks excited by the nonlinear process of Alfv\'en wave greatly contribute to this stellar wind acceleration, which will be explained in \cite{subsequent} in more detail. The magnetic pressure gradient (green line) contributes to supporting the stellar atmosphere and driving the stellar wind within $r\lesssim10r_\star$, but not involved in the further acceleration of stellar wind beyond the distance where the Alfv\'en wave amplitude (orange line) reaches a maximum. The magnetic tension force decelerates the stellar wind against the acceleration by the centrifugal force (blue line). In Figure \ref{fig:figure03.eps}(b), the energy fluxes are normalized by $F_{\rm mass}GM_\star/r_\star$, where $F_{\rm mass}$ is the mass flux and $F_{\rm mass}GM_\star/r_\star\approx5\times10^3r_\star^2/(fr^2)$ erg cm$^{-2}$ s$^{-1}$. It is confirmed that $F_A$, $F_g$, and $F_c$ determine the energy balance at the coronal height $r-r_\star\sim0.1r_\star$, while in the distance ($\gtrsim 10r_\star$), the kinetic energy flux of the stellar wind ($F_{\rm kin}$) dominates the total energy flux. By defining $L_{A,\rm co}$, $L_{g,\rm co}$, $L_{c,\rm co}$ as the energy luminosities $F_AA$, $F_gA$, $F_cA$ at $r=1.1r_\star$ and $L_{\rm kin,wind}$ as $F_{\rm kin}A$ at $r=100r_\star$, the energy conservation along the magnetic flux tube is approximately expressed as:
\begin{equation}
  L_{A,\rm co}\approx L_{\rm kin,wind}-L_{g,\rm co}-L_{c,\rm co}.
  \label{eq:energy_conservation_app}
\end{equation}
The subscript $_{\rm co}$ represents the physical quantities at $r=r_{\rm co}=1.1r_\star$. The above relation shows that the Alfv\'en wave energy flux is converted to the wind's energy loss $L_{\rm kin,wind}-L_{g,\rm co}$ and the coronal heating loss $-L_{c,\rm co}$. Note that $L_{\rm kin,wind}-L_{g,\rm co}=\dot{M}(v_{\rm wind}^2+v_{\rm esc\star}^2)/2$, where $v_{\rm wind}=v_x(r=100r_\star)$ is the wind velocity and $\dot{M}=\rho v_xA$ is the mass loss rate.

\section{STELLAR CORONAE AND WINDS FROM THE SUN TO M DWARFS}

\begin{figure*}
  \begin{center}
    \epsscale{1.} 
    \plotone{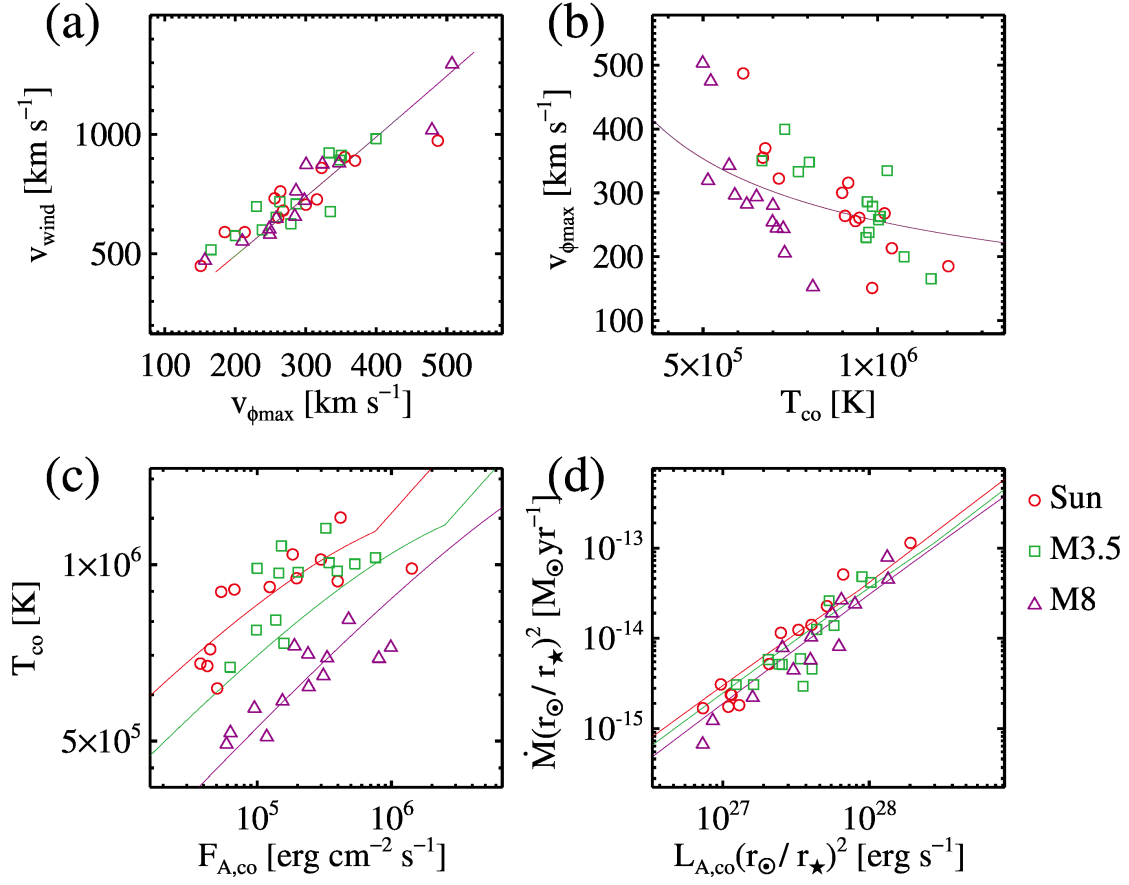}
    \caption{The characteristics of stellar atmospheres and winds obtained from the numerical parameter survey. The symbols represent the simulation results. The parameter survey for each star is carried out about the velocity amplitude on the photosphere (within $(0.04-0.6)\times$ adiabatic sound speeds on the photosphere) and the chromospheric magnetic field strength (within $(0.002-0.05)\times$ photospheric magnetic field strengths). The solid lines correspond to the prediction curves of our semi-empirical method. Panel (a): stellar wind velocity ($v_{\rm wind}$) vs the maximum amplitude of Alfv\'en wave in the stellar wind ($v_{\phi\max}$). Panel (b): $v_{\phi\max}$ vs the coronal temperature ($T_{\rm co}$). Panel (c): $T_{\rm co}$ vs the transmitted Poynting flux into the corona ($F_{A,\rm co}$). Panel (d): Alfv\'en wave luminosity in the corona ($L_{A,\rm co}$) vs stellar wind's mass loss rate ($\dot{M}$).}
    \label{fig:figure04.eps}
  \end{center}
\end{figure*}

Numerical parameter survey about the Sun and M dwarfs reveals the diversity of stellar wind velocity ($v_{\rm wind}$) and coronal temperature ($T_{\rm co}$). Figure \ref{fig:figure04.eps} illustrates the general trends of such characteristics of stellar atmospheres and winds. In Figure \ref{fig:figure04.eps}(a), $v_{\rm wind}$ are plotted as a function of the maximum amplitude of Alfv\'en wave in the stellar wind ($v_{\phi\max}$). The tight correlation between them is accounted for, because $v_{\phi\max}$ well represents the strength of slow shocks which drive the stellar winds. Alfv\'en wave tends to be more amplified in the stellar wind when $T_{\rm co}$ is cooler (Figure \ref{fig:figure04.eps}(b)). Figure \ref{fig:figure04.eps}(c) shows that $T_{\rm co}$ increases with the transmitted Poynting flux into the corona ($F_{A,\rm co}$), but M dwarfs' $T_{\rm co}$ are systematically cooler than that of the Sun for a given $F_{A,\rm co}$. Finally, it is confirmed that the wind's mass loss rates ($\dot{M}$) are well correlated with the energy luminosity of Alfv\'en wave ($L_{A,\rm co}$), as shown in Figure \ref{fig:figure04.eps}(d).

\section{Semi-empirical method to predict the characteristics of stellar atmosphere and wind}

In order to comprehend the physical mechanisms causing the relationships presented in Figure \ref{fig:figure04.eps}, we developed a semi-empirical method to calculate $v_{\rm wind}$, $v_{\phi\max}$ and $T_{\rm co}$ as functions of given effective temperature ($T_{\rm eff}$) and Alfv\'en wave luminosity on the stellar photosphere ($L_{A,\rm ph}$). The derivation of them is briefly summarized in Appendix \ref{sec:semi-empirical} and will be described in \cite{subsequent}. The solid lines in Figure \ref{fig:figure04.eps} are the prediction curves of our semi-empirical method. As shown in Figure \ref{fig:figure04.eps}, the positive or negative correlations among the physical quantities are correctly reproduced by our method, although the simulation results remain scattering around the prediction curves within a factor of $\sim2$. This means that the following scenario which is employed in our semi-empirical method can account for the relationships shown in Figure \ref{fig:figure04.eps}, in qualitative and somewhat quantitative manner.

According to our semi-empirical method, the thinner atmosphere of M dwarf is characterized with increase in the temperature gradient in the corona ($({\rm grad}T)_{\rm co}$) for a given $T_{\rm co}$. The larger $({\rm grad}T)_{\rm co}$ is, the cooler $T_{\rm co}$ is for a given $L_{A,\rm ph}$, so that the energy balance is satisfied between Poynting flux and heat conduction flux. The cooler $T_{\rm co}$ of M dwarf results in lower plasma $\beta$ of stellar wind, in which the amplification of Alfv\'en wave is promoted. The larger amplitude of Alfv\'en wave is associated with the stronger slow shocks which contribute to faster stellar wind of M dwarf. The faster $v_{\rm wind}$ and much smaller surface area of M dwarf lead to much smaller $\dot{M}$ of M dwarf's wind.

By using the established semi-empirical method, we can predict the general trends of $v_{\rm wind}$, $T_{\rm co}$, $\dot{M}$, with respect to $T_{\rm eff}$ and $L_{A,\rm ph}$, as illustrated in Figure \ref{fig:figure05.eps} (see Appendix \ref{sec:semi-empirical}). The open circles in this figure represent the samples of our parameter survey discussed in this letter, about each of which several chromospheric magnetic field strengths are tested. The thick dashed line corresponds to the fiducial $L_{A,\rm ph}$ as a function of $T_{\rm eff}$ which is calculated from the photospheric magnetic field, filling factor of open magnetic flux, and the velocity fluctuation of the convective motion. The thick dash-dotted line corresponds to the largest $L_{A,\rm ph}$ obtained by assuming the convective velocity reaches the sound speed on the photosphere. The thin dashed line represents $L_{A,\rm ph}$ as a function of $T_{\rm eff}$ which results in $v_{\rm wind}=v_{\rm esc\star}$. Along the thick dashed line, it is seen that stellar wind velocity ($v_{\rm wind}$) and coronal temperature ($T_{\rm co}$) are faster and cooler with decreasing $T_{\rm eff}$, and that the mass loss rate ($\dot{M}$) of M-dwarfs' winds are much smaller than the solar wind's value.

\section{Discussion}
Our wind's mass loss rates of M dwarfs are typically smaller than reported by the previous global 3D stellar wind modelings using AWSoM. $\dot{M}$ of M8 type star in this study is no more than $6.9\times10^{-17}$ $M_\odot$ yr$^{-1}$ while \cite{2017ApJ...843L..33G} and \cite{2018PNAS..115..260D} show $3\times10^{-14}$ $M_\odot$ yr$^{-1}$ and $4.1\times10^{-15}$ $M_\odot$ yr$^{-1}$ for TRAPPIST-1 (M8), respectively. $\dot{M}$ of Proxima Centauri (M5.5) by \cite{2016ApJ...833L...4G} and EV Lac (M3.5) by \cite{2014ApJ...790...57C} are $1.5\times10^{-14}$ $M_\odot$ yr$^{-1}$ and $3\times10^{-14}$ $M_\odot$ yr$^{-1}$, respectively, which are $10-100$ times higher than reproduced in our simulation. These much larger mass loss rates probably originate in their inner boundary conditions, corresponding to the top of stellar chromosphere. In particular, our simulation does not validate their estimation of Alfv\'en wave energy injection which is sometimes based on the widely used assumption of the constant ``Poynting-flux-to-field ratio'' \citep{2013ApJ...764...23S}. It is impossible for the above 3D modelings to reproduce our results because they are unable to consider the Alfv\'en wave dissipation and reflection from the stellar photosphere to the top of chromosphere more self-consistently with the present computational resources.

\cite{2011ApJ...741...54C} estimated that $\dot{M}$ of EV Lac (M3.5) is three orders of magnitude smaller than our simulation results about M3.5 type star. This is because they assumed much smaller Poynting flux on the photosphere compared to our simulation. The scaling law about $\dot{M}$ proposed by \cite{2018PASJ...70...34S} also predicts 10-100 times smaller than our estimation. They performed the numerical simulations similar to our study but the low-mass stars with $M_\star\geq 0.6M_\odot$ are considered. According to their analysis, Alfv\'en wave transmissivity into the corona strongly depends on the stellar effective temperature ($\propto T_{\rm eff}^{13/2}$), which possibly leads to the underestimation of $\dot{M}$ for cool dwarfs. Finally, we point out that the assumption of $v_{\rm wind}=v_{\rm esc\star}$ used in both \cite{2011ApJ...741...54C} and \cite{2018PASJ...70...34S} misleadingly implies that $\dot{M}$ depends on $v_{\rm esc\star}$.

Observational measurements of M dwarf's stellar wind are still much challenging. In order to quantify the stellar wind's properties observationally, \citet{2005ApJ...628L.143W} investigated the absorption signatures in stellar Ly$\alpha$ spectra, leading to the estimation of $\dot{M}\sim2\times10^{-14}$ $M_\odot$ yr$^{-1}$ for EV Lac (M3.5). They also suggested an upper limit of Proxima Centauri's $\dot{M}\sim4\times10^{-15}$ $M_\odot$ yr$^{-1}$. \cite{2016A&A...591A.121B} and \cite{2017MNRAS.470.4026V} deduced $\dot{M}$ of GJ 436 (M2.5) around $(0.45-2.5)\times10^{-15}$ $M_\odot$ yr$^{-1}$, by analyzing the transmission spectra of Ly$\alpha$ of GJ 436 b (a warm Neptune). While the observed $\dot{M}$ of GJ 436 and the upper limit on $\dot{M}$ of Proxima Centauri is not inconsistent with our results, the observed $\dot{M}$ of EV Lac is much higher than the simulated value. \cite{2011ApJ...741...54C} argued that the coronal mass ejection (CME) is possibly related to the observed high mass loss rate of EV Lac. To clarify what causes the discrepancy between the observed and simulated $\dot{M}$, further self-consistent modeling is needed for the stellar wind and astrosphere.

\clearpage

\appendix
\section{Basic Equations}
\label{sec:basic_eq}
The basic equations in the axial symmetric magnetic flux tube are written as follows:

\begin{equation}
  {\partial\rho\over\partial t}+{1\over A}{\partial\over\partial x}(\rho v_xA)=0\label{eq:mass},
\end{equation}
\begin{align}
  {\partial\over\partial t}&\left({p\over\gamma-1}+{1\over2}\rho v^2+{B^2\over8\pi}\right)\nonumber\\
  +&{1\over A}{\partial\over\partial x}\left[A\left\{\left({\gamma p\over\gamma-1}+{\rho v^2\over2}+{B_\phi^2\over4\pi}\right)v_x-{B_x\over4\pi}(B_\phi v_\phi)\right\}\right]\nonumber\\
  &=\rho v_x{\partial\over\partial x}\left({GM_\star\over r}\right)-{1\over A}{\partial\over\partial x}(AF_c)-Q_{\mbox{\scriptsize rad}},
  \label{eq:ene}
\end{align}
\begin{align}
  {\partial(\rho v_x)\over\partial t}+&{\partial p\over\partial x}+{1\over A}{\partial\over\partial x}\left\{\left(\rho v_x^2+{B_\phi^2\over8\pi}\right)A\right\}\nonumber\\
  &-\rho v_\phi^2{\partial\ln\sqrt{A}\over\partial x}-\rho{\partial\over\partial x}\left({GM_\star\over r}\right)=0,
  \label{eq:mom_pol}
\end{align}
\begin{equation}
  {\partial(\rho v_\phi)\over\partial t}+{1\over A\sqrt{A}}
  {\partial\over\partial x}\left\{A\sqrt{A}\left(\rho v_xv_\phi-{B_xB_\phi\over4\pi}\right)\right\}=0,
  \label{eq:mom_tor}
\end{equation}
\begin{equation}
  {\partial B_\phi\over\partial t}+{1\over \sqrt{A}}{\partial\over\partial x}\Big(\sqrt{A}(v_xB_\phi-v_\phi B_x)\Big)=0,
  \label{eq:mag_tor}
\end{equation}
\begin{equation}
  {dx\over dr}=\sqrt{1+\left({d\sqrt{A}\over dr}\right)^2}.
\end{equation}
Here, $\gamma$ represents the specific heat ratio and is set to 5/3 in this study. $F_c$ and $Q_{\rm rad}$ are the heat conduction flux and radiative cooling term, respectively. $r$ is the distance from the center of the Sun. $A$ is the cross section of the flux tube (i.e., $B_xA=\rm const.$), and is related to $r$ through the filling factor $f$ as $A(r)=4\pi r^2f(r)$. $f$ determines the geometry of the flux tube. The functions for $F_c$, $Q_{\rm rad}$, and $f(r)$ are similar to those used in \cite{2020ApJ...900..120S}.

\section{Semi-empirical Method for Stellar Coronae and Winds}
\label{sec:semi-empirical}

The derivation of our semi-empirical method is briefly summarized in this appendix. More detailed discussion will appear in our subsequent paper \citep{subsequent}.

In section \ref{sec:results_analyses}, the stellar wind velocity ($v_{\rm wind}$) is determined according to the integral of equation of motion (Equation \ref{eq:bernoulli}).
\begin{equation}
  v_{\rm wind}^2=\Delta_p+\Delta _{p_B}+\Delta_c+\Delta_t+\Delta_g,
  \label{eq:app_eq_0}
\end{equation}
where $v_{\rm wind}$ is the stellar wind velocity at $r=100r_\star$, and $\Delta_p=-2\int^{100r_\star}_{r_\star}\left\langle{1\over\rho}{\partial p\over\partial x}\right\rangle dx$, $\Delta_{p_B}=-2\int^{100r_\star}_{r_\star}\left\langle{1\over\rho}{\partial\over\partial x}\left({B_\phi^2\over8\pi}\right)\right\rangle dx$, $\Delta_c=2\int^{100r_\star}_{r_\star}\left\langle v_\phi^2\right\rangle{\partial\ln\sqrt{A}\over\partial x}dx$, $\Delta_t=-2\int^{100r_\star}_{r_\star}\left\langle{B_\phi^2\over4\pi\rho}\right\rangle{\partial\ln\sqrt{A}\over\partial x}dx$, and $\Delta_g=-v_{\rm esc\star}^2\left(1-{1\over 100}\right)$. They are characterized by the maximum amplitude of Alfv\'en wave in the stellar wind ($v_{\phi\max}$) as follows:
\begin{equation}
  \Delta_p+\Delta_g=a_{1,1}\tilde{v}_{\phi\max}^{k_{1,1}},\ \ \ \ \ 
  \Delta_c=a_{1,2}\tilde{v}_{\phi\max}^{k_{1,2}},\ \ \ \ \
  \Delta_t=-a_{1,3}\tilde{v}_{\phi\max}^{k_{1,3}},\ \ \ \ \
  \Delta_{p_B}=a_{1,4}|\tilde{\Delta}_c+\tilde{\Delta}_t|^{k_{1,4}}.
  \label{eq:app_eq_1}
\end{equation}
where $\tilde{v}_{\phi\max}=v_{\phi\max}/($300 km s$^{-1}$), $\tilde{\Delta}_c=\Delta_c/(319^2$ (km s$^{-1})^2$), $\tilde{\Delta}_t=\Delta_t/(319^2$ (km s$^{-1})^2$). The coefficients ($a_{1,1}$, $a_{1,2}$, $a_{1,3}$, $a_{1,4}$) and power-law indices ($k_{1,1}$, $k_{1,2}$, $k_{1,3}$, $k_{1,4}$) are determined based on our simulation results;
\[
  a_{1,1}=653^2,\ \ \ \ \
  a_{1,2}=585^2,\ \ \ \ \
  a_{1,3}=666^2,\ \ \ \ \
  a_{1,4}=472^2,
\]
in unit of (km s$^{-1})^2$.
\[
  k_{1,1}=2.31,\ \ \ \ \
  k_{1,2}=2.04,\ \ \ \ \
  k_{1,3}=2.12,\ \ \ \ \
  k_{1,4}=0.682.
\]

$v_{\phi\max}$ is negatively correlated with the plasma $\beta$ at the position where Alfv\'en wave amplitude reaches the maximum ($\beta_{\phi\max}$).
\begin{equation}
  v_{\phi\max}=a_2\beta_{\phi\max}^{-k_2},
  \label{eq:app_eq_2}
\end{equation}
where $a_2=286$ km s$^{-1}$ and $k_2=0.171$.

$\beta_{\phi\max}$ is determined by the coronal temperature $T_{\rm co}$ and $v_{\rm wind}$.
\begin{equation}
  \beta_{\phi\max}=a_3\left({\tilde{T}_{\rm co}\over\tilde{v}_{\rm wind}}\right)^{k_3},
  \label{eq:app_eq_3}
\end{equation}
where $\tilde{T}_{\rm co}=T_{\rm co}/(10^6$ K), $\tilde{v}_{\rm wind}=v_{\rm wind}/(600$ km s$^{-1})$. $a_3=2.09\times10^{-2}$ and $k_3=1.85$.

The coronal temperature ($T_{\rm co}$) is determined by the balance between heat conduction flux and the transmitted Poynting flux into the corona, according to the energy conservation law (Equation (\ref{eq:energy_conservation_app})). This is similar to the analytical models of quiescent and flaring coronal loops \citep{1978ApJ...220..643R,1998ApJ...494L.113Y}.
Hereafter, we discuss the following equation which is obtained by dividing the both sides of Equation (\ref{eq:energy_conservation_app}) with $L_{A,\rm co}$.
\begin{equation}
  \alpha_{c/A}=1-\alpha_{{\rm wind}/A}(1+v_{\rm esc\star}^2/v_{\rm wind}^2),
  \label{eq:alpha_c_A}
\end{equation}
where $\alpha_{c/A}$ and $\alpha_{{\rm wind}/A}$ represent the energy conversion efficiency from $L_{A,\rm co}$ to $L_{c,\rm co}$ and $L_{\rm kin,wind}$ (i.e., $L_{c,\rm co}=-\alpha_{c/A}L_{A,\rm co}$ and $L_{\rm kin,wind}=\alpha_{{\rm wind}/A}L_{A,\rm co}$). Note that when $v_{\rm wind}<v_{\rm esc\star}$, $\alpha_{c/A}$ is often quenched to zero which means the approximation for Equations (\ref{eq:energy_conservation_app}) and (\ref{eq:alpha_c_A}) become invalid. To avoid this problem, we assumed the monotonic increase in $L_{c,\rm co}$ with $L_{A,\rm co}$. i.e., $\partial\ln\alpha_{c/A}/\partial\ln L_{A,\rm co}>-1$.

We also confirmed that the coefficient $\alpha_{{\rm wind}/A}$ is almost invariant in our parameter survey about the stars, chromospheric magnetic field strengths, and energy inputs from the photosphere, namely $\alpha_{{\rm wind}/A}=0.442\pm0.166$. Therefore, $\alpha_{{\rm wind}/A}$ is assumed to be constant in this study. It should be noted that, however, $\alpha_{{\rm wind}/A}$ possibly depends on the filling factor of open flux tube ($f_{\rm ph}$), which is beyond our present parameter survey.

By defining the spatial scale of expanding magnetic flux tube ($l_B$) as below, the coronal temperature ($T_{\rm co}$) is estimated as Equation (\ref{eq:app_eq_4}).
  \begin{equation}
    l_B=\int^{\overline{B}/B_{x,\rm co}}_1\left|{d\ln B_x\over dx}\right|^{-1}d\left(B_x\over B_{x,\rm co}\right)
    \label{eq:app_eq_lb}
  \end{equation}
  \begin{equation}
  T_{\rm co}=a_4\left[\left\{1-\alpha_{{\rm wind}/A}\left(1+{v_{\rm esc\star}^2\over v_{\rm wind}^2}\right)\right\}\tilde{F}_{A,\rm co}\tilde{l}_B\right]^{k_4},
  \label{eq:app_eq_4}
  \end{equation}
  where $\overline{B}$ and $B_{x,\rm co}$ are the magnetic field strengths in the chromosphere and corona. $\tilde{F}_{A,\rm co}=F_{A,\rm co}/(10^5$ erg cm$^{-2}$ s$^{-1})$, $\tilde{l}_B=l_B/r_\odot$. $a_4=1.62\times10^6$ K, $k_4=0.256$. Note that $l_B$ is determined only by the assumed geometry of magnetic flux tube.

Finally, $L_{A,\rm co}$ should be expressed as a product of $L_{A,\rm ph}$ which is the Alfv\'en wave luminosity on the stellar photosphere and the transmissivity of Alfv\'en wave from the photosphere to corona ($\alpha_{\rm co/ph}$, i.e., $L_{A,\rm co}=\alpha_{\rm co/ph}L_{A,\rm ph}$). The dissipation and reflection of Alfv\'en wave in the stellar chromosphere could reduce $\alpha_{\rm co/ph}$.

$\alpha_{\rm co/ph}$ is well described by the Alfv\'en travel time from the photosphere to the corona ($\tau_{A,\rm co}$), especially the normalized one by the typical wave frequency of Alfv\'en wave ($\nu_A$). We interpreted $\nu_A$ with the acoustic cutoff frequency of stellar photosphere ($\nu_{\rm ac}$), and found:
\begin{equation}
  \alpha_{\rm co/ph}=a_{5,1}(\tau_{A,\rm co}\nu_{\rm ac}/a_{5,2})^{k_5},
  \label{eq:app_eq_5}
\end{equation}
where $a_{5,1}=2.41\times10^{-2}$ and $a_{5,2}=1.04$. $k_5=1.25$ when $\tau_{A,\rm co}\nu_{\rm ac}<1.04$, and otherwise, $k_5=-1.10$. $\tau_{A,\rm co}\nu_{\rm ac}$ is empirically expressed as a function of $g_\star$, chromospheric magnetic field strength ($\overline{B}$), and velocity amplitude on the photosphere ($v_{\rm ph}$):
\begin{equation}
  \tau_{A,\rm co}\nu_{\rm ac}=a_6\tilde{g}_\star^{k_{6,1}}\left(\overline{B}\over B_{\rm ph}\right)^{-k_{6,2}}\left(v_{\rm ph}\over c_{s,\rm ph}\right)^{k_{6,3}},
  \label{eq:app_eq_6}
\end{equation}
where $c_{s,\rm ph}$ and $B_{\rm ph}$ are the adiabatic sound speed and magnetic field strength on the photosphere. $a_6=0.921$, $k_{6,1}=0.240$, $k_{6,2}=0.408$, and $k_{6,3}=0.697$.

Based on Equations (\ref{eq:app_eq_lb})$-$(\ref{eq:app_eq_6}), $T_{\rm co}$ is obtained as a function of $v_{\rm wind}$ and $L_{A,\rm ph}$ (or $v_{\rm ph}$) by specifying the basic parameters ($g_\star$, $B_{\rm ph}$, $\overline{B}$, $c_{s,\rm ph}$, $v_{\rm esc\star}$, $\alpha_{{\rm wind}/A}$, $l_B$). These parameters can be related to the stellar effective temperature $T_{\rm eff}$ by limiting our discussion to the main-sequence stars' atmospheres and winds. On the other hand, Equations (\ref{eq:app_eq_0})$-$(\ref{eq:app_eq_3}) show that $v_{\rm wind}$ should be determined implicitly when $T_{\rm co}$ is given. By using some iterative method, therefore, $T_{\rm co}$ and $v_{\rm wind}$ are calculated for given $L_{A,\rm ph}$ and $T_{\rm eff}$. From the obtained $v_{\rm wind}$ and the definition of $L_{\rm kin,wind}$, the mass loss rate of stellar wind ($\dot{M}$) is expressed as below:
\begin{equation}
  \dot{M}=2\alpha_{{\rm wind}/A}\alpha_{\rm co/ph}{L_{A,\rm ph}\over v_{\rm wind}^2}
\end{equation}

We will present \cite{subsequent} to explain the derivation of the above coefficients ($a_{1,1}$, $a_{1,2}$, $a_{1,3}$, $a_{1,4}$, $a_2$, $a_3$, $a_4$, $a_{5,1}$, $a_{5,2}$, $a_6$) and power-law indices ($k_{1,1}$, $k_{1,2}$, $k_{1,3}$, $k_{1,4}$, $k_2$, $k_3$, $k_4$, $k_5$, $k_{6,1}$, $k_{6,2}$, $k_{6,3}$) with more simulation results for some M dwarfs (M0, M5, M5.5).

\begin{figure*}
  \begin{center}
    \epsscale{1} 
    \plotone{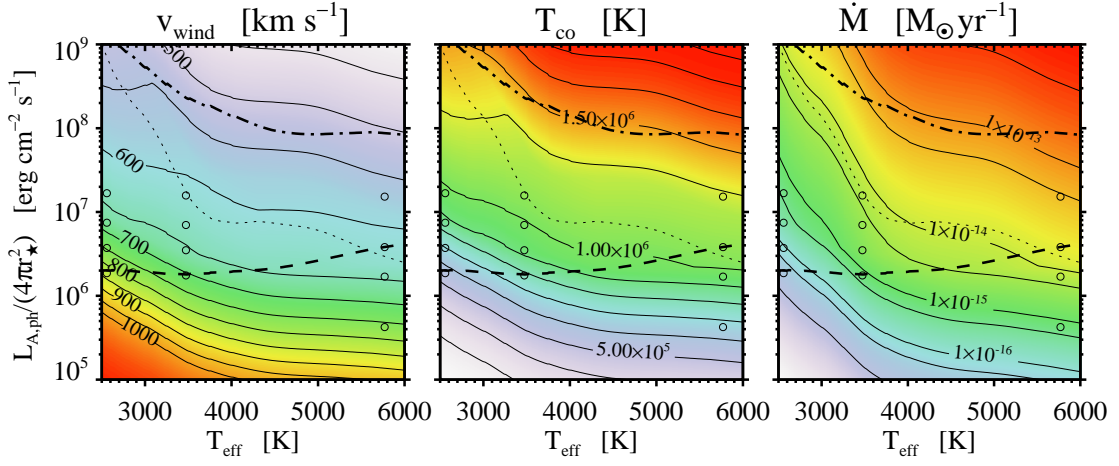}
    \caption{The general trends of stellar wind velocity ($v_{\rm wind}$), coronal temperature ($T_{\rm co}$), wind's mass loss rate ($\dot{M}$), with respect to the effective temperature ($T_{\rm eff}$) and Alfv\'en wave luminosity on the photosphere ($L_{A,\rm ph}$), which is predicted by our semi-empirical method. The open circles in this figure represent part of samples in our parameter survey. The thick dashed line corresponds to the fiducial $L_{A,\rm ph}$ as a function of $T_{\rm eff}$. The thick dash-dotted line corresponds to the largest $L_{A,\rm ph}$ obtained by assuming the convective velocity reaches the sound speed on the photosphere. The thin dashed line represents $L_{A,\rm ph}$ as a function of $T_{\rm eff}$ which results in $v_{\rm wind}=v_{\rm esc\star}$.}
    \label{fig:figure05.eps}
  \end{center}
\end{figure*}

\clearpage

\acknowledgments
T.S. was supported by JSPS KAKENHI Grant Number JP18J12677. A part of this study was carried out by using the computational resources of the Center for Integrated Data Science, Institute for Space-Earth Environmental Research, Nagoya University through the joint research program, XC40 at YITP in Kyoto University, and Cray XC50 at Center for Computational Astrophysics, National Astronomical Observatory of Japan. Numerical analyses were partly carried out on analysis servers at the Center for Computational Astrophysics, National Astronomical Observatory of Japan.




\begin{thebibliography}{}
\bibitem[Alvarado-G{\'o}mez et al.(2020)]{2020ApJ...895...47A} Alvarado-G{\'o}mez, J.~D., Drake, J.~J., Fraschetti, F., et al.\ 2020, \apj, 895, 47. doi:10.3847/1538-4357/ab88a3
\bibitem[Bourrier et al.(2016)]{2016A&A...591A.121B} Bourrier, V., Lecavelier des Etangs, A., Ehrenreich, D., et al.\ 2016, \aap, 591, A121
\bibitem[Cohen et al.(2014)]{2014ApJ...790...57C} Cohen, O., Drake, J.~J., Glocer, A., et al.\ 2014, \apj, 790, 57
\bibitem[Cranmer \& Saar(2011)]{2011ApJ...741...54C} Cranmer, S.~R., \& Saar, S.~H.\ 2011, \apj, 741, 54
\bibitem[Dong et al.(2018)]{2018PNAS..115..260D} Dong, C., Jin, M., Lingam, M., et al.\ 2018, Proceedings of the National Academy of Science, 115, 260
\bibitem[Freedman et al.(2014)]{2014ApJS..214...25F} Freedman, R.~S., Lustig-Yaeger, J., Fortney, J.~J., et al.\ 2014, \apjs, 214, 25
\bibitem[Garraffo et al.(2016)]{2016ApJ...833L...4G} Garraffo, C., Drake, J.~J., \& Cohen, O.\ 2016, \apjl, 833, L4
\bibitem[Garraffo et al.(2017)]{2017ApJ...843L..33G} Garraffo, C., Drake, J.~J., Cohen, O., Alvarado-G{\'o}mez, J.~D., \& Moschou, S.~P.\ 2017, \apjl, 843, L33
\bibitem[Gillon et al.(2016)]{2016Natur.533..221G} Gillon, M., Jehin, E., Lederer, S.~M., et al.\ 2016, \nat, 533, 221
\bibitem[Hollweg et al.(1982)]{1982SoPh...75...35H} Hollweg, J.~V., Jackson, S., \& Galloway, D.\ 1982, \solphys, 75, 35
\bibitem[Khodachenko et al.(2007)]{2007AsBio...7..167K} Khodachenko, M.~L., Ribas, I., Lammer, H., et al.\ 2007, Astrobiology, 7, 167
\bibitem[Kudoh \& Shibata(1999)]{1999ApJ...514..493K} Kudoh, T., \& Shibata, K.\ 1999, \apj, 514, 493
\bibitem[Lammer et al.(2007)]{2007AsBio...7..185L} Lammer, H., Lichtenegger, H.~I.~M., Kulikov, Y.~N., et al.\ 2007, Astrobiology, 7, 185 
\bibitem[Linsky(2019)]{2019LNP...955.....L} Linsky, J.\ 2019, Lecture Notes in Physics, Berlin Springer Verlag
\bibitem[Ludwig et al.(1999)]{1999AA...346..111L} Ludwig, H.-G., Freytag, B., \& Steffen, M.\ 1999, \aap, 346, 111
\bibitem[Ludwig et al.(2002)]{2002AA...395...99L} Ludwig, H.-G., Allard, F., \& Hauschildt, P.~H.\ 2002, \aap, 395, 99
\bibitem[Magic et al.(2015)]{2015AA...573A..89M} Magic, Z., Weiss, A., \& Asplund, M.\ 2015, \aap, 573, A89
\bibitem[Maldonado et al.(2015)]{2015AA...577A.132M} Maldonado, J., Affer, L., Micela, G., et al.\ 2015, \aap, 577, A132
\bibitem[Matsumoto \& Shibata(2010)]{2010ApJ...710.1857M} Matsumoto, T., \& Shibata, K.\ 2010, \apj, 710, 1857
\bibitem[Matsumoto \& Suzuki(2012)]{2012ApJ...749....8M} Matsumoto, T., \& Suzuki, T.~K.\ 2012, \apj, 749, 8
\bibitem[Matsumoto \& Suzuki(2014)]{2014MNRAS.440..971M} Matsumoto, T., \& Suzuki, T.~K.\ 2014, \mnras, 440, 971
\bibitem[Matsumoto(2020)]{2020MNRAS.tmp.3336M} Matsumoto, T.\ 2020, \mnras. doi:10.1093/mnras/staa3533
\bibitem[Mesquita \& Vidotto(2020)]{2020MNRAS.494.1297M} Mesquita, A.~L., \& Vidotto, A.~A.\ 2020, \mnras, 494, 1297
\bibitem[Meyer et al.(2012)]{2012MNRAS.422.2102M} Meyer, C.~D., Balsara, D.~S., \& Aslam, T.~D.\ 2012, \mnras, 422, 2102
\bibitem[Miyoshi \& Kusano(2005)]{2005JCoPh.208..315M} Miyoshi, T., \& Kusano, K.\ 2005, Journal of Computational Physics, 208, 315
\bibitem[Rosner et al.(1978)]{1978ApJ...220..643R} Rosner, R., Tucker, W.~H., \& Vaiana, G.~S.\ 1978, \apj, 220, 643
\bibitem[Sakaue \& Shibata(2020)]{2020ApJ...900..120S} Sakaue, T. \& Shibata, K.\ 2020, \apj, 900, 120
\bibitem[Sakaue \& Shibata(2021)]{subsequent} Sakaue, T. \& Shibata, K.\ 2021, to be submitted. 
\bibitem[Scalo et al.(2007)]{2007AsBio...7...85S} Scalo, J., Kaltenegger, L., Segura, A.~G., et al.\ 2007, Astrobiology, 7, 85
\bibitem[Seager(2013)]{2013Sci...340..577S} Seager, S.\ 2013, Science, 340, 577
\bibitem[Shoda et al.(2018)]{2018ApJ...853..190S} Shoda, M., Yokoyama, T., \& Suzuki, T.~K.\ 2018, \apj, 853, 190
\bibitem[Shoda et al.(2019)]{2019ApJ...880L...2S} Shoda, M., Suzuki, T.~K., Asgari-Targhi, M., et al.\ 2019, \apjl, 880, L2
\bibitem[Shoda et al.(2020)]{2020ApJ...896..123S} Shoda, M., Suzuki, T.~K., Matt, S.~P., et al.\ 2020, \apj, 896, 123
\bibitem[Shu \& Osher(1988)]{1988JCoPh..77..439S} Shu, C.-W., \& Osher, S.\ 1988, Journal of Computational Physics, 77, 439 
\bibitem[Sokolov et al.(2013)]{2013ApJ...764...23S} Sokolov, I.~V., van der Holst, B., Oran, R., et al.\ 2013, \apj, 764, 23
\bibitem[Suzuki \& Inutsuka(2005)]{2005ApJ...632L..49S} Suzuki, T.~K., \& Inutsuka, S.-i.\ 2005, \apjl, 632, L49
\bibitem[Suzuki \& Inutsuka(2006)]{2006JGRA..111.6101S} Suzuki, T.~K., \& Inutsuka, S.-I.\ 2006, Journal of Geophysical Research (Space Physics), 111, A06101 
\bibitem[Suzuki et al.(2013)]{2013PASJ...65...98S} Suzuki, T.~K., Imada, S., Kataoka, R., et al.\ 2013, \pasj, 65, 98
\bibitem[Suzuki(2018)]{2018PASJ...70...34S} Suzuki, T.~K.\ 2018, \pasj, 70, 34
\bibitem[Tarter et al.(2007)]{2007AsBio...7...30T} Tarter, J.~C., Backus, P.~R., Mancinelli, R.~L., et al.\ 2007, Astrobiology, 7, 30
\bibitem[van der Holst et al.(2014)]{2014ApJ...782...81V} van der Holst, B., Sokolov, I.~V., Meng, X., et al.\ 2014, \apj, 782, 81
\bibitem[Velli(1993)]{1993A&A...270..304V} Velli, M.\ 1993, \aap, 270, 304
\bibitem[Vidotto et al.(2014)]{2014MNRAS.438.1162V} Vidotto, A.~A., Jardine, M., Morin, J., et al.\ 2014, \mnras, 438, 1162
\bibitem[Vidotto \& Bourrier(2017)]{2017MNRAS.470.4026V} Vidotto, A.~A., \& Bourrier, V.\ 2017, \mnras, 470, 4026
\bibitem[Wood et al.(2005)]{2005ApJ...628L.143W} Wood, B.~E., M{\"u}ller, H.-R., Zank, G.~P., Linsky, J.~L., \& Redfield, S.\ 2005, \apjl, 628, L143
\bibitem[Yokoyama \& Shibata(1998)]{1998ApJ...494L.113Y} Yokoyama, T. \& Shibata, K.\ 1998, \apjl, 494, L113
\end{thebibliography}
\end{document}